\documentstyle[preprint,aps]{revtex}
\def\be{\begin{equation}}
\def\ee{\end{equation}}
\begin{document}
\draft

\title{Structural transitions in a classical two-dimensional molecule
system}

\author{W. P. Ferreira$^{1}$, G. A. Farias$^{1}$, H. A. Carmona$^{2}$,
and F. M. Peeters$^{3}$}
\address{$^1$Departamento de F\'{\i}sica, Universidade Federal do Cear\'a,
Caixa Postal 6030,\break Campus do Pici, 60455-760 Fortaleza,
Cear\'a, Brazil.\\$^2$Departamento de F\'{\i}sica e Qu\'{\i}mica,
Universidade Estadual do Cear\'a, Av. Paranjana, 1700 Fortaleza,
Cear\'a, Brazil.\\$^3$Departement Natuurkunde, Universiteit
Antwerpen (UIA), Universiteitsplein 1, B-2610 Antwerpen, Belgium}

\date{\today}

\maketitle

\begin{abstract}

The ground state of a classical two-dimensional (2D) system with a
finite number of charges particles, trapped by two positive
impurity charges localized at a distance $(z_0)$ from the 2D plane
and separated from each other by a distance $x_p$ are obtained.
The impurities are allowed to carry more than one positive charge.
This classical system can form a 2D-like classical molecule that
exhibits {\it structural transitions} and {\it spontaneous
symmetry breaking} as function of the separation between the
positive charges before it transforms into two 2D-like classical
atoms. We also observe structural transitions as a function of the
dielectric constant of the substrate which supports the charged
particles, in addition to broken symmetry states and {\it
unbinding of particles.}
\end{abstract}







\narrowtext

\newpage

\section{INTRODUCTION}

Classical charged particles confined in a two-dimensional (2D)
layer has been a subject of interest in the last few
years.\cite{grimes,ando,clark,dubin,chu,jean,aref} It is
well-known that classical 2D electrons form a Wigner
crystal.\cite{grimes,isihara} On the other hand, confined charged
particles in 2D form 2D-like classical atoms \cite{peeters1} which
can exhibit very different melting and structural phase
transitions.\cite{schweigert1} Recently these properties were
studied for a set of charged particles confined by a parabolic and
hard-wall well.\cite{peeters1,schweigert1,peeters2,schweigert2}
They order in a ring structure and a Mendeleev table for these
atomic like structures was constructed. An essentially different
confinement is the one produced by a Coulomb potential. 2D charges
which are confined by this potential also form 2D-like classical
atomic structures but they exhibit very different structural
transitions as compared to the one confined by parabolic
potentials.\cite{farias}

In the present paper we extend our previous work for single
classical atoms \cite{peeters1,farias} and vertically coupled
atoms \cite{partoens} to laterally coupled atoms which form
molecular structures. The system we consider here is composed of
classical charged point particles interacting through a repulsive
Coulomb potential and held together by two remote impurities a
distance $d$ away from the 2D plane where the charged particles
can move. The ground state of the system is calculated by using
the Monte Carlo simulation technique and by minimization
techniques. We show that the ground state of this 2D-like
classical molecule (2DCM) presents structural phase transitions
and if the impurities are separated further apart it is broken
into two 2D-like classical atoms (2DCA). Spontaneous broken
symmetry and evaporation is observed, which depend strongly on the
effective impurity charges and the distance between them.

\section{THEORETICAL MODEL}

 Our model consists of a finite number of particles $N$, each of
them with negative charge  $e = -\mid e \mid$, which move in a 2D
plane. Two fixed impurities, each of them with charge $+Ze$, are
localized a distance $z_o$ from the 2D plane and are separated by
a distance $X_{P}$ along the $x$-axis and are placed inside a
medium with dielectric constant $\epsilon$. All particles interact
through a Coulomb potential (see Fig. 1). The energy for such a
classical system is given by

\begin{equation}
E = \frac{- Ze^{2} }{\epsilon} \sum_{i=1}^{N} \sum_{j=1}^{2}
\frac{1}{\mid\vec{R}_i - \vec{D}_j\mid} +
\frac{e^{2}}{\epsilon{_o}}\sum_{j>i=1}^{N} \frac{1}{\mid\vec{R}_i
- \vec{R}_j\mid}, \label{eq:eq1}
\end{equation}

\noindent where $\vec{R}_i$ is the position of the $i^{\it th}$
negative particle at the $(X,Y)$ plane and $\vec{D}_j =
D_x\vec{e}_x - Z_o\vec{e}_z$ ($D_x=X_p/2$) is the position of the
$j^{\it th}$ positive charge, {\it i.e.} the impurity. To keep the
problem as simple as possible we have neglected in
Eq.~(\ref{eq:eq1}) the effects due to the image charges. Following
our previous work \cite{farias} we will vary the dielectric
constant $\epsilon$ through which it is possible to change the
effective confinement potential in a continuous way. For
convenience, we express the energy in units of $E_o =
e^{2}/{\epsilon_{o} z_0}$ where $\epsilon_{o}$ is the dielectric
constant of vacuum and all the distances are now in units of
$z_0$. This reduces Eq.~(\ref{eq:eq1}) to

\begin{equation}
E^{*}= \frac{- Z^{*}}{\epsilon}
\sum_{i=1}^{N}\sum_{j=1}^{2}\frac{1}{\mid\vec{r}_i -
\vec{d}_j\mid} + \sum_{j>i=1}^{N}\frac{1}{\mid\vec{r}_i -
\vec{r}_j\mid}, \label{eq:eq2}
\end{equation}

\noindent where $E^{*} =E /E_0$, $Z^{*}=Z/\epsilon^{*}$,
$x_p=X_p/z_0$, $\vec{d}_j=\vec{D}_j/z_0$,
$\vec{r}_j=\vec{R}_j/z_0$, $\epsilon^{*}=\epsilon/\epsilon_{o}$.
In order to obtain the minimum energy configuration we perform a
numerical minimization of Eq.~(\ref{eq:eq2}) with respect to the
$2N$ variables (coordinates $(x,y)$ of the negative particles).

\section{RESULTS}

To perform the numerical simulation, we start from a random
distribution of negative particles and obtain the minimum energy
configuration by using standard minimization techniques. Next we
change, {\it e.g.} $x_p$, slightly and take as the initial
configuration the last minimum obtained by our numerical
minimization. To have an independent check we also use the Monte
Carlo simulation technique starting from several different initial
configurations. Different from the results obtained with the 2D
classical atom \cite{farias} we found that bounded configurations
, {\it i.e.}, $r_i < \infty$ $\forall$ $i$, can exist if the total
effective charge $Z_T = 2Z^{*}$ satisfies the condition $Z_T \leq
N -1$ for $N> 1$. In fact, we observed that the condition for
bounded configurations tends to behave as a single 2-D classical
atom, if the distance between the two positive charges goes to
zero. To analyze the molecule-atom transition we fix the value of
the dielectric constant $\epsilon^{*}=1$, we consider a system of
six electrons and put the two positive charges at the points
$(d_{x},0,-1)$ and $(-d_{x},0,-1)$, each with a charge $Z=3$. In
Fig. 2 we show the distance $r_i$ of the six negative particles to
the point $(x,y,z)=(0,0,0)$ in the minimum energy configuration as
a function of the distance between the positive charges $x_{p} = 2
d_{x}$. We clearly observe four distinct regions. In the first
region (I), $x_{p}$ is very small and the negative particles are
arranged in two equilateral triangles, inside each other (see Fig.
3(a)). The two positive charges (their positions are indicated by
the crosses in Fig. 3) are so close to each other that they behave
as one central charge. This configuration is very different from
the one which was found in the case of a parabolic confinement
potential \cite{peeters1} where one particle would be at
(x,y)=(0,0) and the other 5 in a ring around it. At $x_{p} =
0.0884$ the system undergoes a small structural transition to
region (II)(see Fig. 3(b)) which is of first order. Note that
increasing $x_p$ from $x_p=0$ acts as a symmetry breaking field
which breaks the rotational symmetry of the configuration depicted
in Fig. 3(a). With increasing $x_{p}$, the ground state
configuration changes continuously up to $x_p = 1.0074$ where it
undergoes a structural transition and a new configuration appears
in region (III). This transition is characterized by the fact that
the particles change their positions abruptly (compare solid (II)
and open (III) symbols in Fig. 3(c)). With further increase of
$x_{p}$ the system exhibits a new structural transition to region
(IV) at $x_p = 1.8792 $ where the configuration clearly exhibits
now inversion symmetry with respect to the center of the
coordinate system. In region (IV) the negative particles tend to
stay in two triangles, each one close to one of the positive
charges. Up to this region, the system behaves as a classical
molecule. Namely, all distances between the negative particles are
related to the positions of both positive charges.
Finally, for a value $x_{p} \simeq 3.0$ we observe more explicitly
that the negative particles split in two sets, each one correlated
to one positive charge. No structural transition is associated
with this behavior but, up to this point, the system behaves as
two independent 2D classical atoms. This can be seen by the fact
that, up to $x_{p} \simeq 3.0$ the system exhibits rotational
symmetry along the $z$-axis
for each of the two positive charges. Different from the
structural transition from regions (I) up to (IV), the one from 2D
classical molecule to 2D classical atom is not characterized by a
structural transition, i.e. it is continuous. In order to
visualize these stable states, in Fig. 3 we show typical
configurations of the system in each region and in Fig. 4 we plot
the distance of three negative particles related to each positive
charge. As shown in Fig. 4, when $x_p \geq 3.0$ the system behaves
as two independent atoms, with the negative particles in two
equilater triangles and the positive charges in the center of each
one, with energy $E_{single} \rightarrow - 4.899$.
In order to show the order of the  structural transitions we
calculated the first of the energy as a function of $x_{p}$, by
numerically differentiating the energy curve. The results for the
first derivative are presented in Fig. 5. As can be seen the first
derivative is discontinuous at the structural transitions (I
$\rightarrow$ II, II $\rightarrow$ III and III $\rightarrow$ IV).
It is continuous throughout the region where the classical 2D
molecule transits to two 2D classical atoms.

Different structural transitions are observed if we analyze the
behavior of the minimum energy configuration as a function of the
dielectric constant ($\epsilon^{*}$). To this, we consider two
positive charges (each one with $Z = 3$) fixed at the points
$(1,0,-1)$ and $(-1,0,-1)$ and six negative particles which are
free to move in the xy-plane. For $\epsilon^{*} = 1$ we are in
region IV of the above. In Fig. 6 we show the distance of those
six negative particles to the point $(x,y,z)=(0,0,0)$ for the
minimum energy configuration as a function of $\epsilon^{*}$.
Eleven different regions are observed, of which we plot in Fig. 6
only the first ten of them. In region (I),
$1.0<\epsilon^{*}<1.054$, the minimum energy configuration
corresponds to the negative particles arranged in two triangles
each one close to each positive charge (see Fig. 7(a)). When
$\epsilon^{*}$ is increased a structural transition takes place at
$\epsilon^{*} = 1.117$. Up to this value of $\epsilon^{*}$, region
(III), the configuration changes continuously and for
$\epsilon^{*}\rightarrow 1.225$ one of the negative particles
becomes weakly bound and moves to infinity, i.e. it evaporates.
After this value, region (IV), the minimum energy configuration
changes continuously and again presents a structural transition,
at $\epsilon^{*} =1.314$, to a small region (V). This region goes
up to $\epsilon^{*} =1.325$, where a new structural transition
takes place. Now, in the new region, region (VI), another negative
particle starts to become unbound when the value $\epsilon^{*}
=1.492$ is reached. The minimum energy configuration is stable in
the new region (VII) until the value $\epsilon^{*} = 1.909$ and
again changes its structure. After this point, again another
particle starts to move away, region (VIII), and goes to infinity
at $\epsilon^{*} = 2.112$. Up to this point, region (IX), only
three particles remain bound until $\epsilon^{*} =3.000$. In
region(X) $16.2> \epsilon^{*}> 3.000$ only two particles remain,
with each negative particle sitting almost on top of each of the
positive charges in a symmetrical arrangement. Finally, when
$\epsilon^{*} > 16.2$ only one particle remains bounded.
Unexpectedly, this negative particle stays in an asymmetric
position, in this case at a distance $r_i =0.77$. This fact can be
confirmed by taking the minimum of the potential energy. In this
case it presents a minimum at $r_i =0.77$ and a local maximum at
$r_i =0$, which corresponds to a symmetrical arrangement. The
existence of region (X) shows that a bounded configuration can
exist in a region where the condition $Z_T \leq N -1$ for $N> 1$
is not satisfied.

\section{CONCLUSIONS}

The molecule-atom transition and structural transitions were
studied in a 2D model system of negative particles confined by a
Coulomb potential of two positive impurity charges. This system
presents a large variety of structural transitions as a function
of its parameters, the distance between the impurity charges and
the dielectric constant. As observed for a 2D like atom the 2D
like molecule exhibits highly asymmetric ground state
configurations which are related to broken symmetry states.
However, the condition $Z_T \leq N -1$ for  the existence of a
bounded configuration is only satisfied if the positive particles
are very close. Finally, our results show that a finite 2D system
confined by a nonlinear potential presents a much larger wealth of
structural transitions and new states as compared to the ones
confined by parabolic potentials.

\bigskip

\acknowledgments W. P. Ferreira, G. A. Farias, and H. A. Carmona
are supported by the Brazilian Agencies, Funda\c{c}\~ao Cearense
de Amparo \`a Pesquisa (FUNCAP), Brazilian National Research
Council(CNPq) and The Ministry of Planning (FINEP). The work of F.
M. Peeters is supported by the Flemish Science Foundation
(FWO-Vl), the EU-RTN network on 'Surface electrons on mesoscopic
structures' and INTAS.

\vfill\eject

\vfill\eject

\begin{figure}[1] \vglue1truecm
\caption{Schematic representation of the system considered in this
paper.} \label{fig1}
\end{figure}

\begin{figure}[2]
\caption{Distance $r_i$ of the six negative particles to the point
$(x,y,z)=(0,0,0)$ for the minimum energy configuration as a
function of the distance between the two positive particles
$x_{p}$. Structural transitions are indicated by the vertical
dotted lines.}
\label{fig2}
\end{figure}

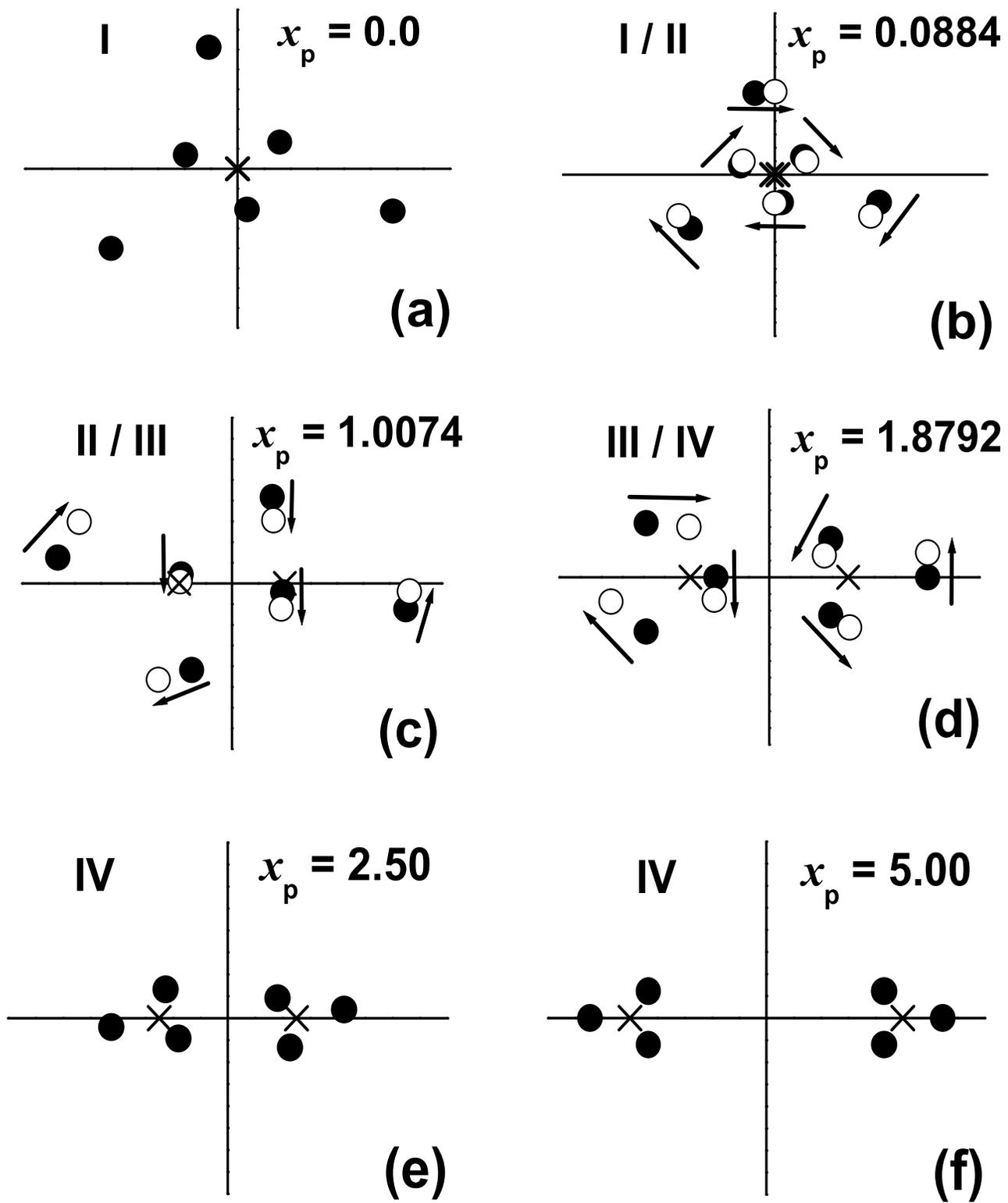
\begin{figure}[3]
\caption{Typical configurations (and transitions) of the system in
the different regions corresponding to Fig. 2.}
\label{fig3}
\end{figure}

\begin{figure}[4]
\caption{Distance of three negative particles related to each
positive charge. Structural transitions are indicated by the
vertical dotted lines.}
\label{fig4}
\end{figure}

\begin{figure}[5]
\caption{The first derivative of the energy as a function of
$x_{p}$. The inset is an enlargement of the small $x_p$ region of
the discontinuity of the first derivative. Structural transitions
are indicated by the vertical dotted lines.}
\label{fig5}
\end{figure}

\begin{figure}[6]
\caption{Distance of six negative particles to the point
$(x,y,z)=(0,0,0)$ in the minimum energy configuration as a
function of the logarithm of the dielectric constant
$\epsilon^{*}$. The dashed lines indicate structural transitions
while the vertical dotted lines correspond to the evaporation of a
particle.}
\label{fig6}
\end{figure}

\begin{figure}[7]
\caption{Typical stable configurations of the ground state in
different regions of $\epsilon^{*}$.}
\label{fig7}
\end{figure}








\vfill\eject

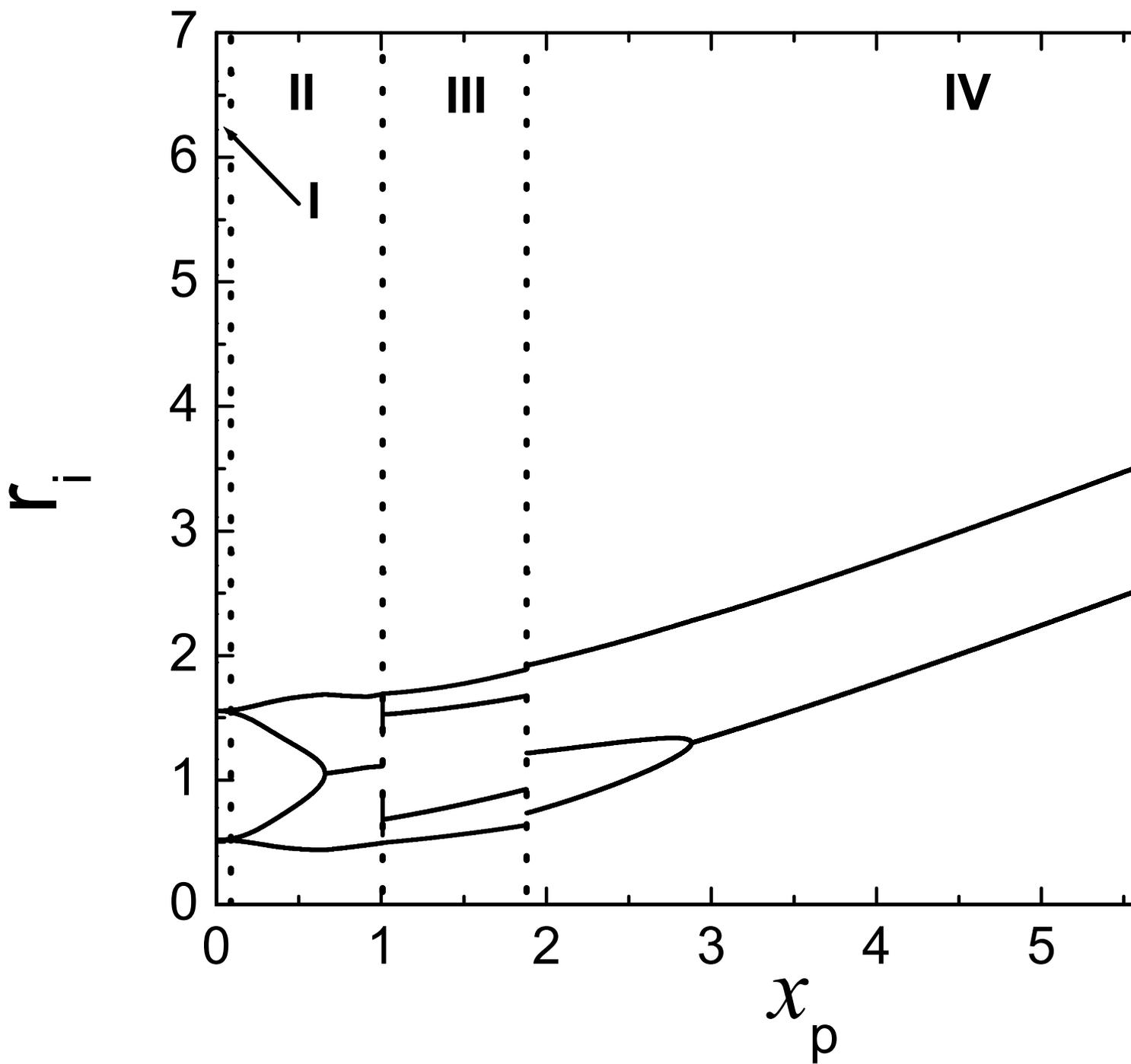
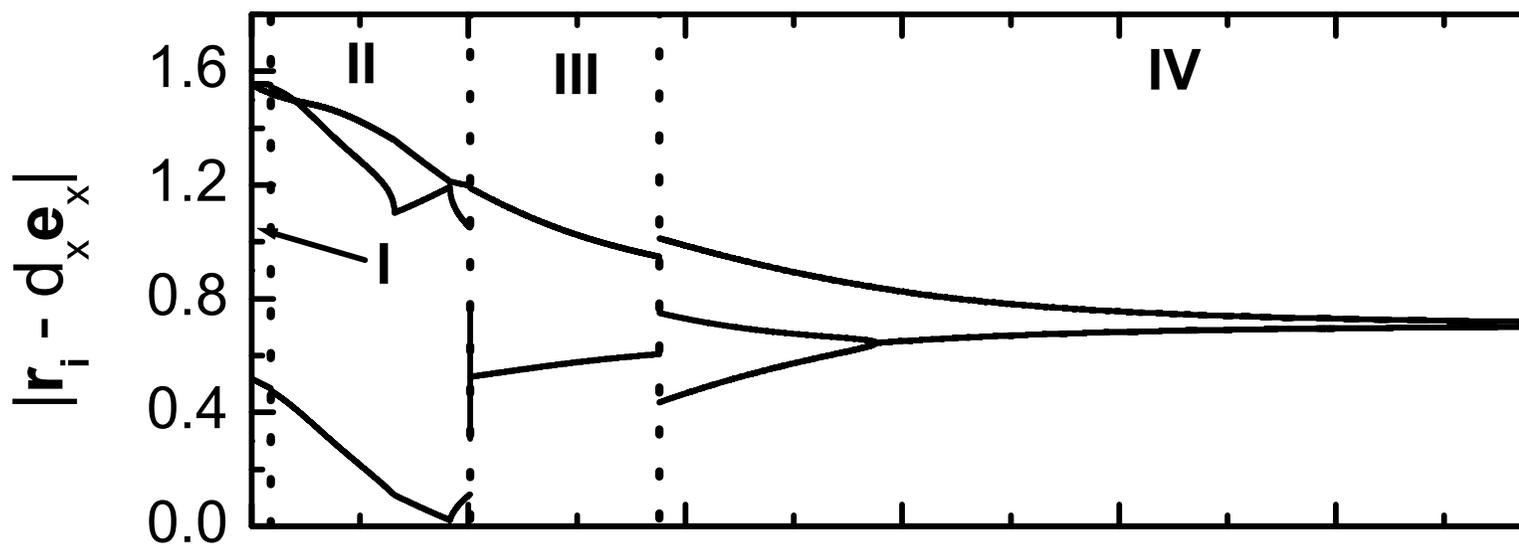
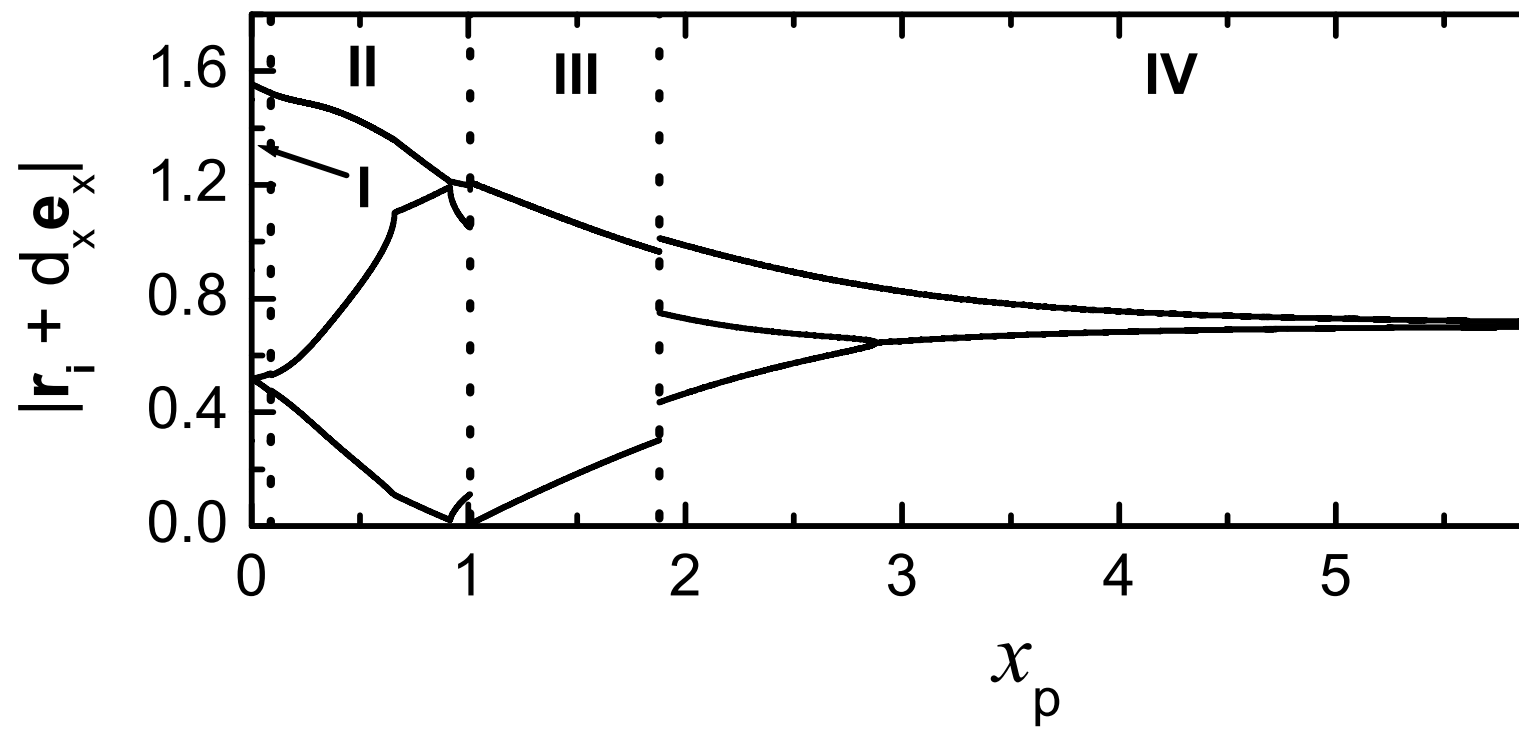
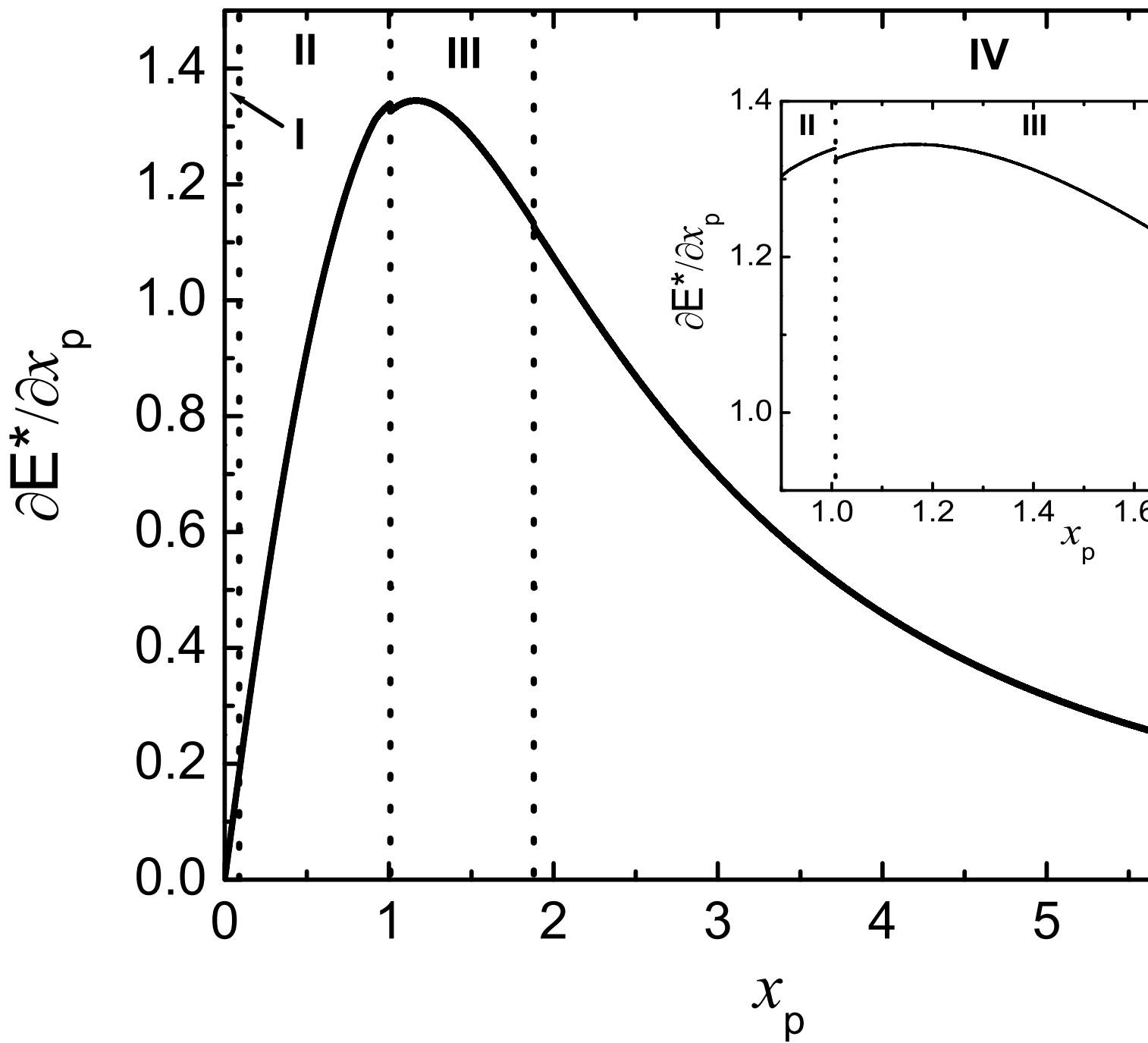
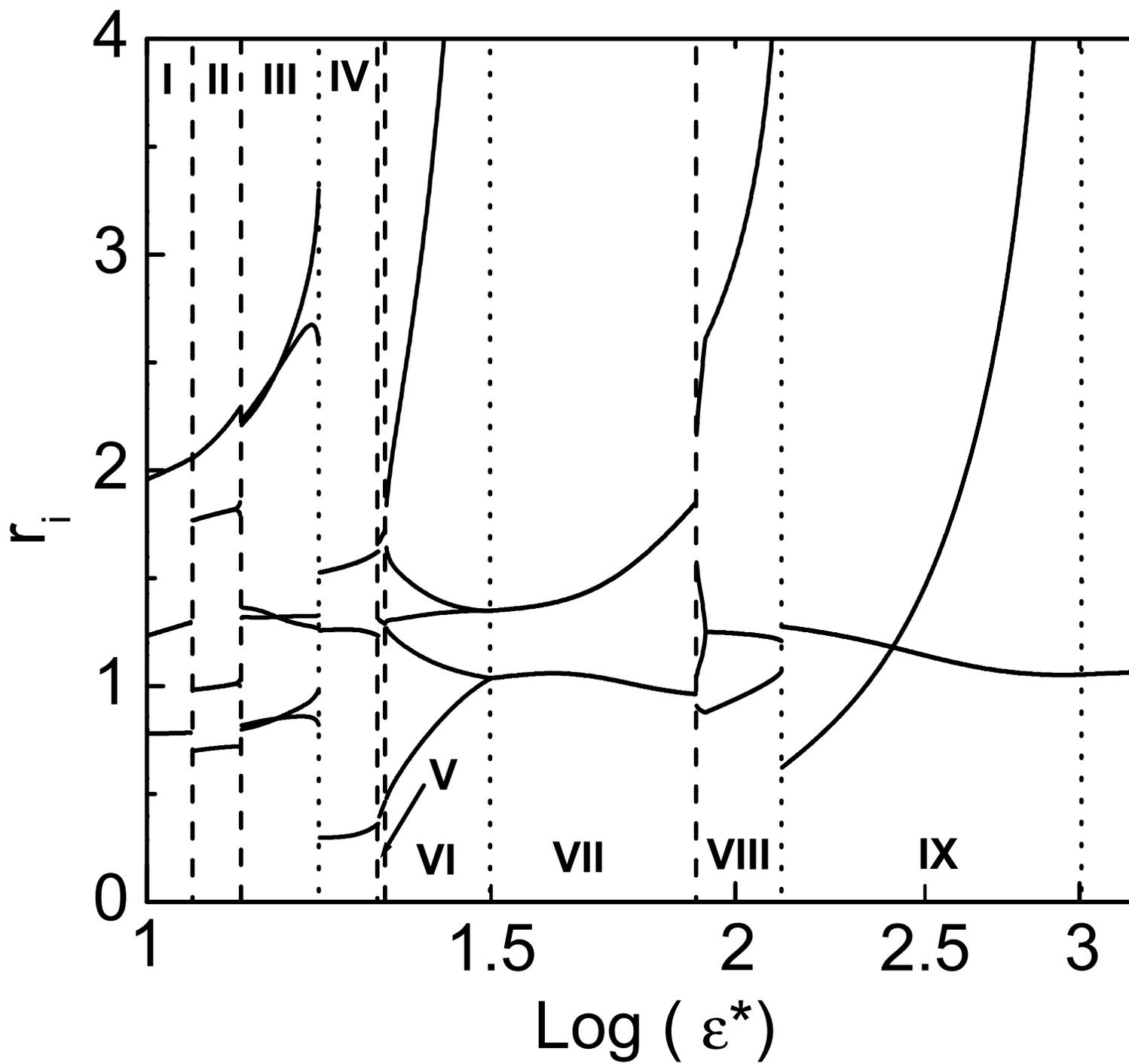
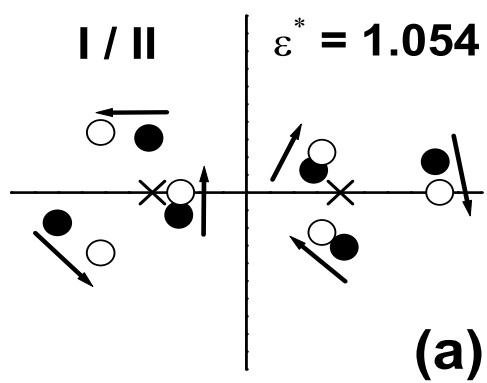
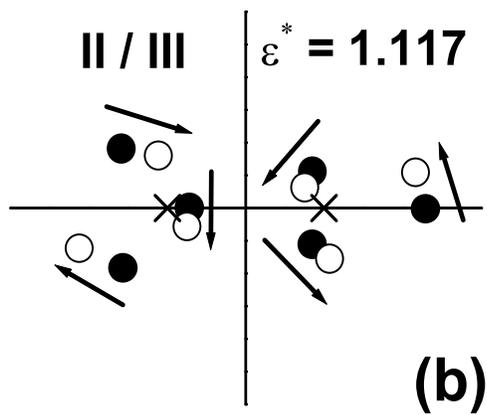
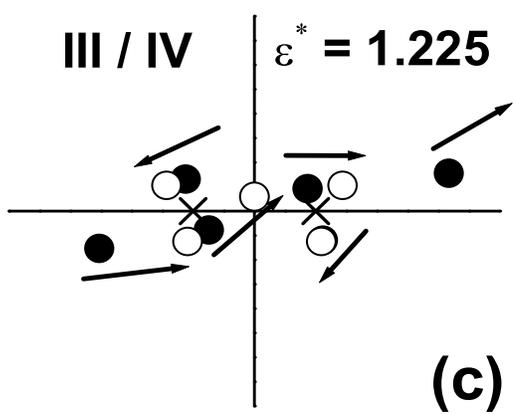
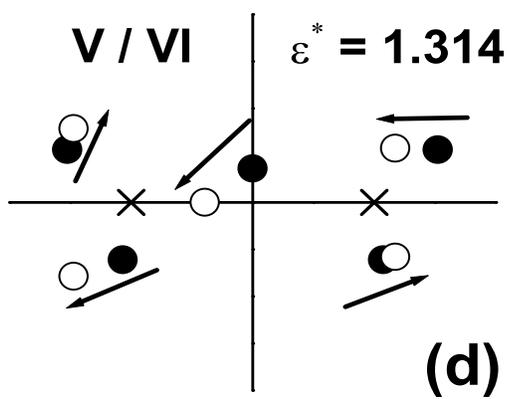
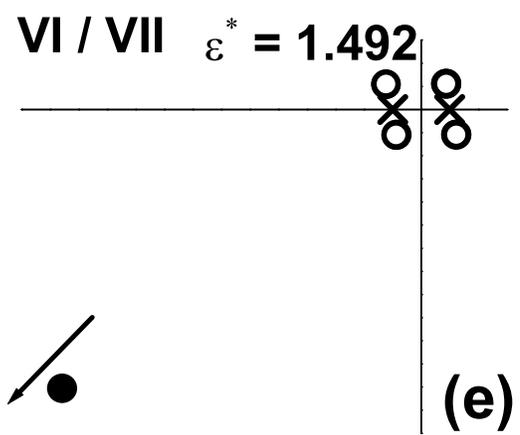
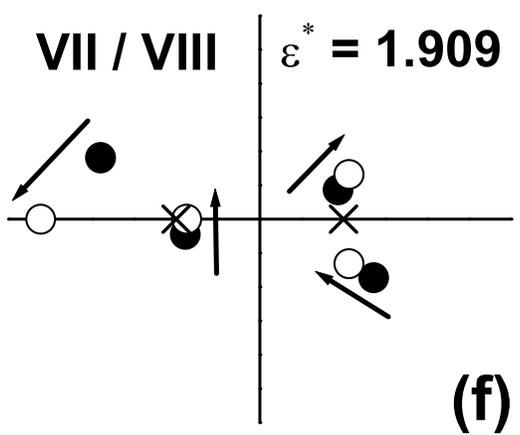
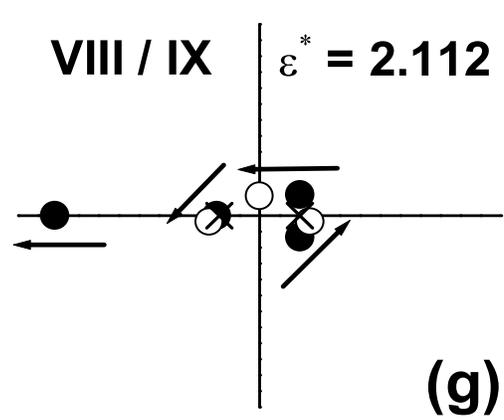
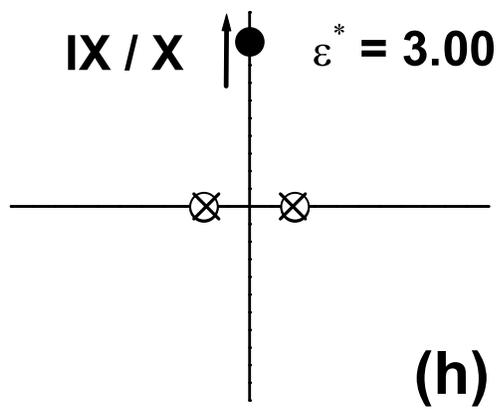
\end{document}